\documentclass[letterpaper,twocolumn,pra,aps,showspace,superscriptaddress,floatfix,amsmath,amssymb]{revtex4-1}
\usepackage[latin1]{inputenc}
\usepackage{bm}
\usepackage{multirow,amssymb,amsbsy,amsmath}
\usepackage{graphicx}
\usepackage{verbatim}
\usepackage{color}
\usepackage{hyperref}
\makeatletter
\usepackage{pifont}
\makeatother

\usepackage{graphicx}% Include figure files
\usepackage{dcolumn}% Align table columns on decimal point
\usepackage{bm}% bold math
\usepackage{ifthen}
\usepackage{booktabs}
\usepackage{nicefrac}
\usepackage{float}

\begin{document}

\title{Experimentally Detecting a Quantum Change Point via Bayesian Inference}
\author{Shang Yu}
\author{Chang-Jiang Huang}
\author{Jian-Shun Tang}
\email{tjs@ustc.edu.cn}
\author{Zhih-Ahn Jia}
\author{Yi-Tao Wang}
\author{Zhi-Jin Ke}
\author{Wei Liu}
\author{Xiao Liu}
\author{Zong-Quan Zhou}
\author{Ze-Di Cheng}
\author{Jin-Shi Xu}
\author{Yu-Chun Wu}
\author{Yuan-Yuan Zhao}
\author{Guo-Yong Xiang}
\email{gyxiang@ustc.edu.cn}
\author{Chuan-Feng Li}
\email{cfli@ustc.edu.cn}
\author{Guang-Can Guo}
\affiliation{CAS Key Laboratory of Quantum Information, University of Science and Technology of China, Hefei, Anhui 230026, China.}
\affiliation{Synergetic Innovation Center of Quantum Information $\&$ Quantum Physics, University of Science and Technology of China, Hefei, Anhui 230026, China.}
\author{Gael Sent\'{\i}s}
\email{gael.sentis@uni-siegen.de}
\affiliation{Naturwissenschaftlich-Technische Fakult\"at, Universit\"at Siegen, 57068 Siegen, Germany.}
\author{Ramon Mu\~{n}oz-Tapia}
\email{Ramon.Munoz@uab.cat}
\affiliation{F\'{i}sica Te\`{o}rica: Informaci\'{o} i Fen\`{o}mens Qu\`antics, Departament de F\'{\i}sica, Universitat Aut\`{o}noma de Barcelona, 08193 Bellaterra (Barcelona), Spain.}

\date{\today}
\begin{abstract}
Detecting a change point is a crucial task in statistics that has been recently extended to the quantum realm. A source state generator that emits a series of single photons in a default state suffers an alteration at some point and starts to emit photons in a mutated state. The problem consists in identifying the point where the change took place. In this work, we consider a learning agent that applies Bayesian inference on experimental data to solve this problem. This learning machine adjusts the measurement over each photon according to the past experimental results and finds the change position in an online fashion. Our results show that the local-detection success probability can be largely improved by using such a machine learning technique. This protocol provides a tool for improvement in many applications where a sequence of identical quantum states is required.
\end{abstract}

\maketitle %\nopagebreak
\bibliographystyle{prsty}
%\section{Introduction}
The change point problem is a crucial concept in statistics~\cite{Pollak1985,Baseville1993,Chen2012} that has been studied in many real-world situations, from stock markets~\cite{Chen1997} to protein folding~\cite{Pirchi2011}. One of the key goals in this field of research is to device procedures that detect the exact point where a sudden change has occurred. This point could indicate, for example, the trigger of a financial crisis or a mis-folded protein step~\cite{physorg}. Many settings can be formulated as a change point problem, however we can understand it in the simplest terms as a Heads or Tails game. Alice sends a series of fair coins to Bob who can toss each of them. She then starts sending  another type of coin and Bob's task is to recognize from his observations the most likely point where the switching occurred.

In recent works, this problem has been extended to the quantum realm~\cite{Akimoto2011,Sentis2016,Sentis2017}. Consider a quantum state generator that is supposed to emit a sequence of photons in a default state, but suffers an uncontrolled alteration at some unknown point (e.g., a rotation of the polarization). Given the sequence of photons, the problem is to determine where the change took place from measurements on the photons. The most general procedure consists in waiting until all the photons have reached the detector and measuring them at the very end~\cite{Sentis2016,physorg}. An optimal global measurement naturally provides the best identification performance, but it is difficult to realize experimentally as it requires a quantum memory to store photons as well as collective quantum operations. On the other hand, it is much more feasible to  measure the state of each photon as soon as it arrives (local measurement). The most basic local procedure consists of a fixed measurement that unambiguously detects a mutated state with some probability. Unfortunately the success probability of this simple procedure is far below the optimal one. In Ref.~\cite{Sentis2016}, a Bayesian Inference procedure that adapts the local measurements according to the previous outcome was also proposed. Such a protocol can be viewed as a Machine Learning (ML) mechanism.

There are many recent works that have proved that a learning machine (either classical~\cite{Sentis2016,Wiebe2014,Carleo2017,Wang2017,Deng2017,Chng2017,Mavadia2017,Torlai2017} or quantum~\cite{Harrow2009,Wiebe2012,Lloyd2014,Rebentrost2014}) can provide an efficient route to quantum characterization, verification and validation. We have already benefited from using conventional machine learning to solve many quantum problems~\cite{Carleo2017,Wang2017,Deng2017,Chng2017,Mavadia2017,Torlai2017,Wan2016,Lu2017}. In particular, Bayesian inference as a classic technique in the ML domain has been discussed in Ref.~\cite{Biamonte2017,Wang2017,Wiebe2014,Paesani2017,Wiebe2016} and applied to quantum tasks in many cases, such as quantum Hamiltonian learning~\cite{Wang2017,Wiebe2014}, phase estimation~\cite{Paesani2017,Wiebe2016,Xiang2011,Berry2001,Berry2007}, and several others~\cite{Burnier2013,Javanainen2015}(spectral function reconstruction and Josephson oscillation characterization).

In this work, we experimentally detect a quantum change point in a sequence of photons using Bayesian inference (BI) and basic local (BL) strategies~\cite{Sentis2016}. We compare the success probabilities between these two methods and with respect to the (theoretical) optimal global strategy. For the BI strategy, we build a learning agent (a programmed computer) to guess the change point position. Once the photon arrives, we detect the result and update the priors for each hypothesis. Then the agent decides which measurement basis is used for the next photon. In other words, the detection basis is being learnt as the protocol proceeds and a refined guess is provided at each step. The output guess is given after the last measurement. In contrast, the measurement basis of the BL strategy is fixed during the whole experiment, and the guessed change position is determined by the first conclusive detection of a mutated state. The BI and BL strategies are described in detail in Appendix~\ref{app:A}. In order to compare the performances of these methods, we conduct a total of 1000 experiments for a certain overlap between the default and mutated state (50 times for one possible change point, and there are 20 possible change points in our situation). Our results show that the learning agent provides a significant advantage over the BL detection and a performance very close to the optimal (global) one.

\begin{figure}[tb]
\centering
\includegraphics[width=0.5\textwidth]{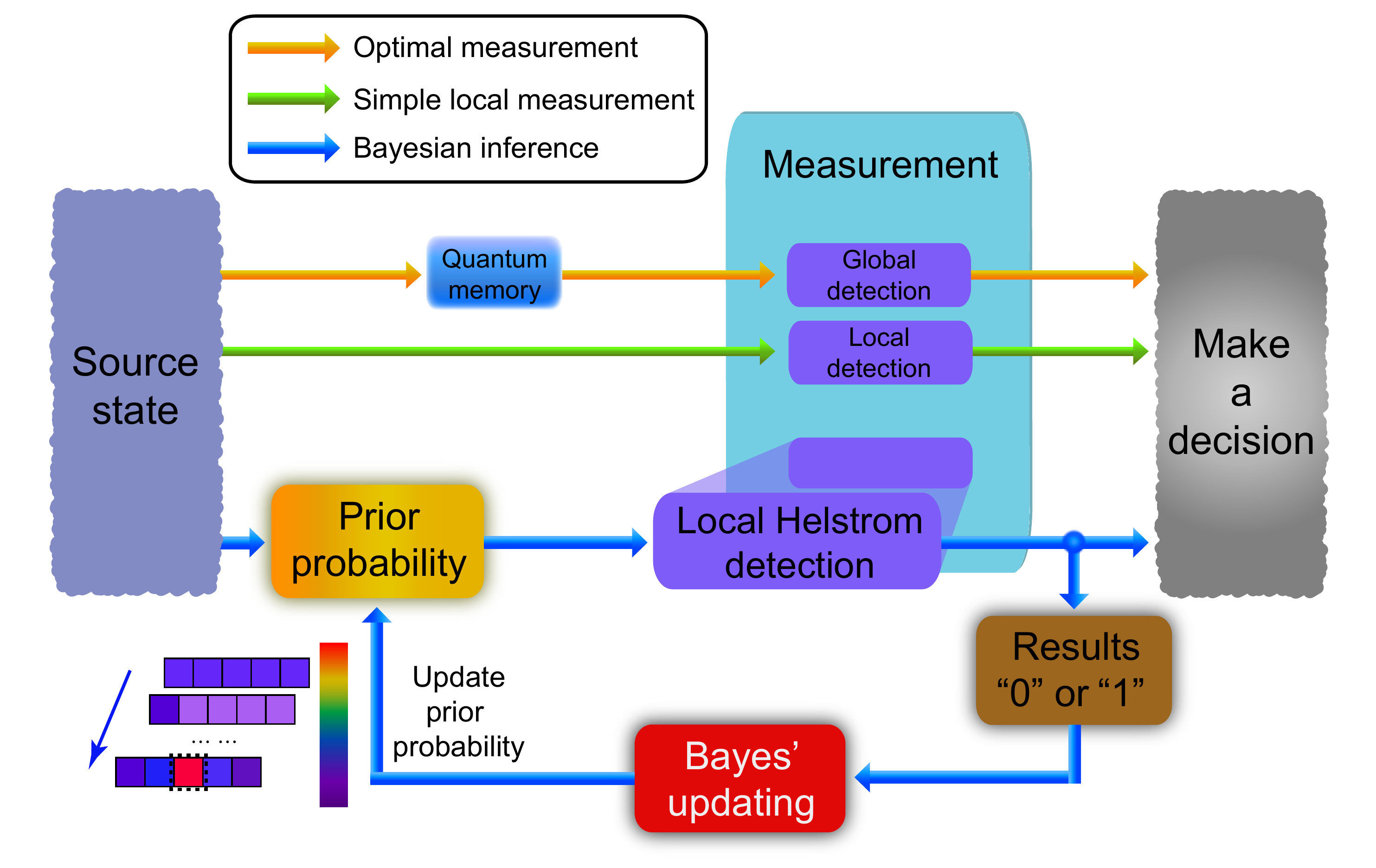}
\caption{\label{Fig1} Schematic diagram of the three different detection protocols. The optimal measurement is constituted by four steps: the source state preparation, a quantum memory for storing the each photon, global detection, and decision making. For the simple local strategy, such as basic local (BL) detection, the quantum memory model is removed and global detection is replaced by a local detection. The measurement base is fixed in BL method, but is changed in the Bayesian inference (BI) strategy. The Helstrom measurement in this method is decided by the prior probability obtained from last step~(see Appendix~\ref{app:A}). Meanwhile, the priors are also updated according the measurement results using the Bayesian update rule. The change point is determined by choosing the position with the largest prior probability after the last step.}
\end{figure}

\begin{figure}[tb]
\centering
\includegraphics[width=0.5\textwidth]{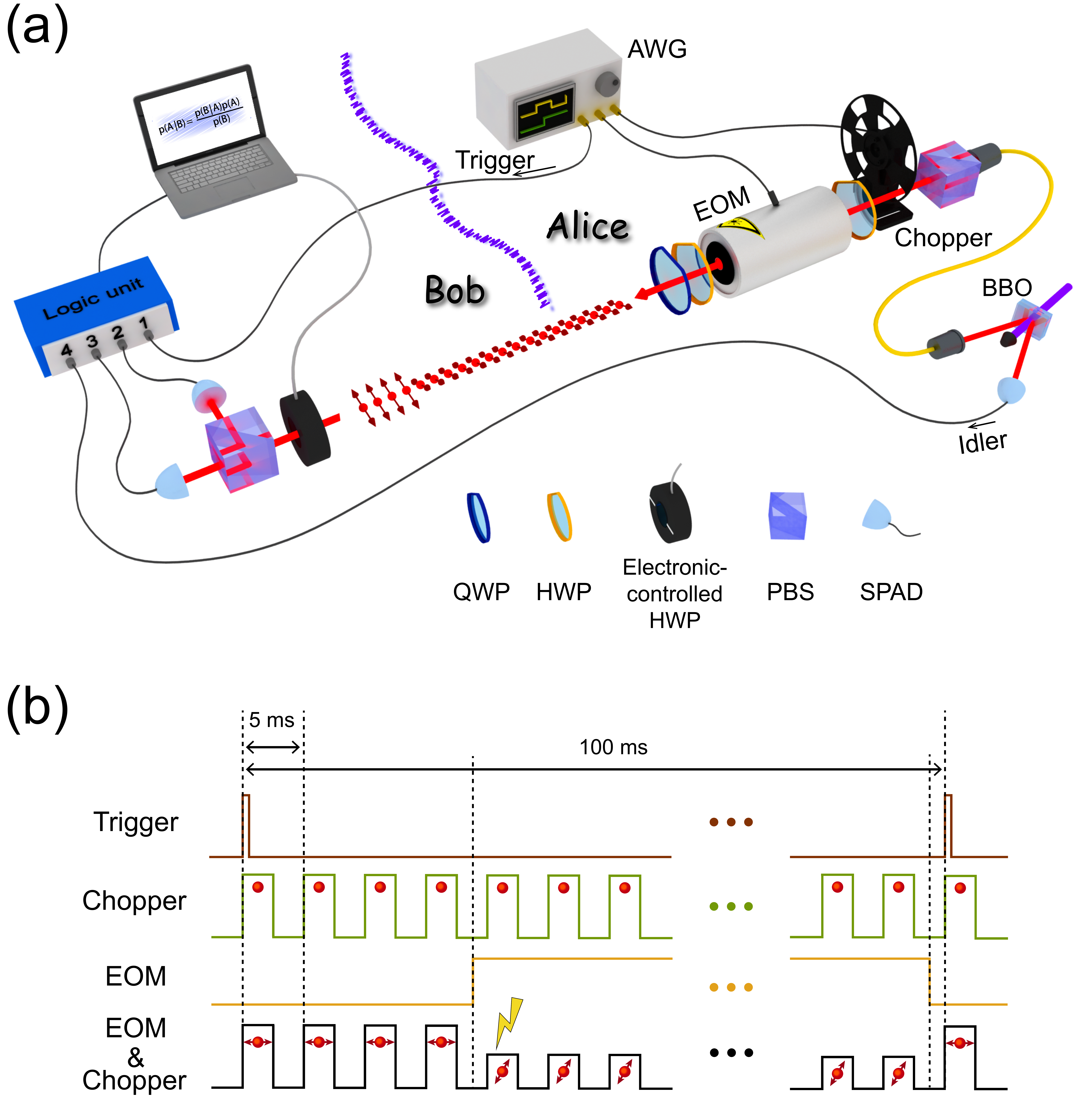}
\caption{\label{Fig2} (a) Experimental setup for the Bayesian inference-based detection measurement. This setup consists of two main parts: source state preparation at Alice side and the detection agent controlled by Bob. The heralded single photon is generated by the SPDC process in BBO crystal. The photons then pass through a PBS to produce the default states $\vert H \rangle$ and an EOM after that creates the change point at some certain place in the sequence. The chopper between them divides the photons into $n$ equal discrete time bins, and we postselect the first effective event from the $i$th time bin as the $i$th particle in source state~(see Appendix~\ref{app:B} for details). The measurement device consists of an electronic-controlled HWP, a PBS, and a classical learning agent (personal computer). The measurement basis realized by the PBS and the electronic-controlled HWP can respond rapidly when the learning agent calculates the new basis from the latest priors. (b) The time sequence diagram. The first line is the trigger signal (generated by AWG), which has a 100 ms interval. The second line denotes the time bins created by the chopper. Each time bin has a 2.5 ms width and the red balls represent signal photons postselected by the logic unit. The third and fourth lines are the EOM signal and the folding signal by EOM and chopper, respectively. In the picture, we show a case where the change point is set at the $5$th photon, and the arrows on the red balls denote the photon polarization.}
\end{figure}

The logical diagram of the detection methods is depicted in Fig. 1.
If the change occurs at the $k$th photon, the source state consisted by $n$ particles can be expressed as
\begin{equation}
\vert \Psi_{k} \rangle= \vert H \rangle^{\otimes k-1}\vert \phi \rangle^{\otimes n-k+1},
\end{equation}
here, $\vert H\rangle$ is the default state and $\vert \phi \rangle=c\vert H\rangle+s\vert V\rangle$ denotes the mutated state (without loss of generality, we set $c$ to be real and positive).
The BI approach contains the following processes~\cite{Sentis2016}: first, the learning agent calculates an updated prior probability (starting from a uniform prior probability at the first step); second, the agent performs a Helstrom measurement~\cite{Helstrom1976} which is decided according to the prior distribution;  third, depending on the outcome, the priors are renewed according to Bayes updating rule. The procedure is repeated until the last photon is measured and the decision is  given by the hypothesis with highest updated prior. The detailed calculations are given in the Appendix~\ref{app:A}.

The experimental setup is sketched in Fig. 2(a), which can be recognized as two parts: the source state generator, which is controlled by Alice; and a classical learning agent, which is owned by Bob. Here the heralded single photon is generated by a type-II spontaneous parametric down-conversion (SPDC) process in BBO crystals. While Alice begins to prepare the source state, the arbitrary waveform generator (AWG) will send a trigger signal (at 100 ms intervals, the first line of the time sequence diagram, as shown in Fig. 2(b)) to Bob and inform him the detection task should begin. Then, she lets the photons pass through a polarization beam splitter (PBS) to obtain the default states $\vert H \rangle$. A followed chopper driven by the AWG is applied to separate the series of single photons into $n$ equal time bins (see the second line in Fig. 2(b)), which determines the number of particles in the source state (the logic unit postselects the first effective event in each bin to be the particles in the source state, see Appendix~\ref{app:B}).
In our experiment, we set $n=20$ (the corresponding chopper period is 5 ms). An electro-optic modulator (EOM), in conjunction with half-wave plates (HWPs) and a quarter-wave plate (QWP), are used to change the default state $\vert H \rangle$ into a mutated state $\vert \phi \rangle$ after some certain point. During the experiment, Alice can decide at which point ($k$) the mutation occurs by controlling the relative time delay of the signals output from the AWG (see the third and fourth lines in Fig. 2(b)). After these procedures, the source state $\vert \Psi_{k} \rangle$ is prepared.

Once the trigger signal is received, Bob starts his measurement. At the first step, he tunes the electronic-controlled HWP to the basis calculated by the uniform prior distribution. Then, a two-outcome measurement is performed on the first photon and detected by two single photon avalanche diodes (SPADs). After that, the result (``0'' or ``1'')~\cite{results} is sent to the computer to calculate the priors $\eta_{k}^{(2)}$ according to the Bayes' updating rules (the details are shown in the Appendix~\ref{app:A}). A new measurement basis then can be determined for the next step. Until the last measurement is finished, Bob produces a guess $\hat{k}$ that maximizes $\eta_{k}^{(21)}$ for the change point.

An example of a single learning process is shown in Fig. 3(a). We choose a random value of the overlap, $c^2=0.604$, and a change point at position $k=5$.  The height of the columns in each row represent the priors for every hypothesis $k=1,\ldots, 20$ after each measurement step. The initial priors are set to be uniform as Bob is assumed not to have prior knowledge  about the position of the change point.   As the measurements proceed, the  Bayesian inference method is able to learn the right position and to correct previous wrong guesses and mistakes caused by experimental noise: we can find that the prior distribution begins to converge on the correct point after the sixth step, although an incorrect value occurred at the third step (the highest column occurs at $k = 2$, which is not the correct change point position). As can be readily seen in Fig. 3(a), the highest updated prior probability at the end of the process $\hat{k}=\arg \max_k \{\eta_k^{(21)}\} $ is precisely $\hat{k}=5$.

\begin{figure}[tb]
\centering
\includegraphics[width=0.48\textwidth]{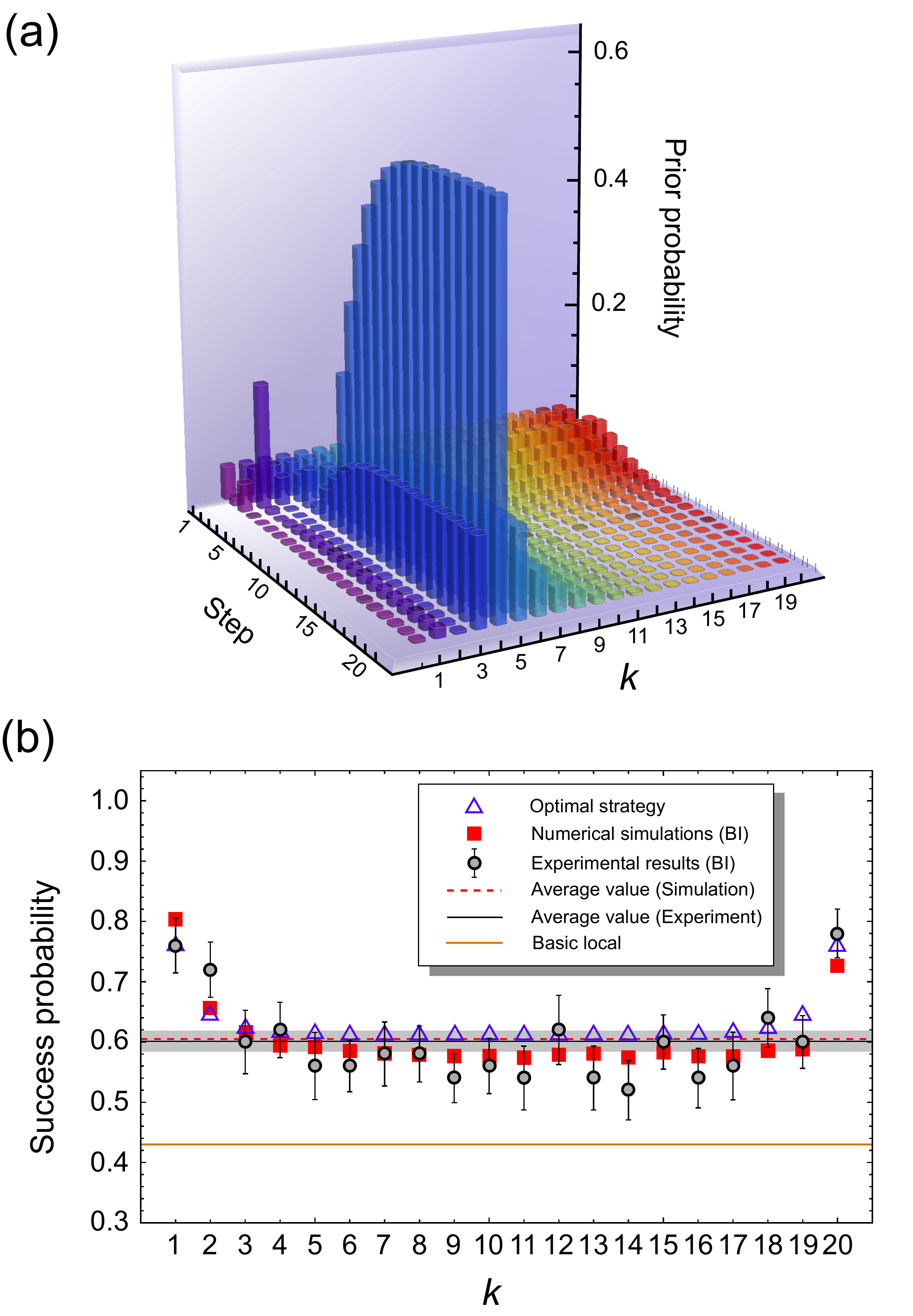}
\caption{\label{Fig3} (a) The results of prior probabilities at each step when $c^{2}=0.604$ and $k=5$. From the first step to the last ($21$st) step, the priors fluctuate depend on the measurement results and begin to converge at the sixth step. At the last step, the highest column is located at $k=5$, which corresponds to a prior probability of 0.569. This result corresponds to the correct change point in this experiment. (b) The relationship between success probability and change point position. Here, Alice sets $c^{2}=0.604$, but varies each change position. The purple triangles correspond to the success probabilities for the optimal (global) strategy~\cite{Sentis2016}, the red squares are the numerical simulation results for the BI strategy and the circles are the actual experimental results. The simulation average value of the success probabilities are denoted by the dash red line. The black solid line and the gray region are the average values and associated errors obtained from experiments, respectively. It is clear that the probabilities derived from BI protocol are all beyond the BL one, denoted as the solid orange line.}
\end{figure}

Next, we analyze the detection performance for each position of the change point. We repeat the above experiment 50 times to gather statistics and compute the success probabilities for each source state $\vert\Psi_k\rangle$, $k=1,2, \ldots, 20$  (we fix the overlap to be same as before, i.e., $c^2= 0.604$). The conditional success probabilities as a function of $k$  are shown in Fig. 3(b). The red squares represent the numerical results of a Monte Carlo simulation of the experiment, and the actual experimental data are shown as circles. Notice the good agreement between both values that coincide within the experimental error bars. We find that most change positions ($k = 3 \sim 19$) have constant success probabilities of approximately 0.58  without large fluctuations. We also observe that the first two  and the last position  can be better detected than the rest (with a success probability larger than 0.7). This is an expected result, as change points occurring toward the beginning or end of the sequence are more distinguishable from its neighbors than those occurring at the bulk of the sequence. For comparison, we also show the optimal (global) conditional probabilities with purple triangles. One can also appreciate from Fig. 3(b) that there is a small but systematic difference between the local BI and the global protocol for most  values of $k$. However, at the end points, this difference disappears and the BI protocol performs almost optimally~\cite{position}. Furthermore, the success probabilities produced by the classical learning agent all surpass the BL method (the solid orange line) including the errors. This advantage remains even for large $n$. More numerical simulation results are given in the Appendix~\ref{app:C}. Here, the error-bars are obtained by 100 Monte-Carlo simulations (we use the same number of simulations below).

\begin{figure}[tb]
\centering
\includegraphics[width=0.48\textwidth]{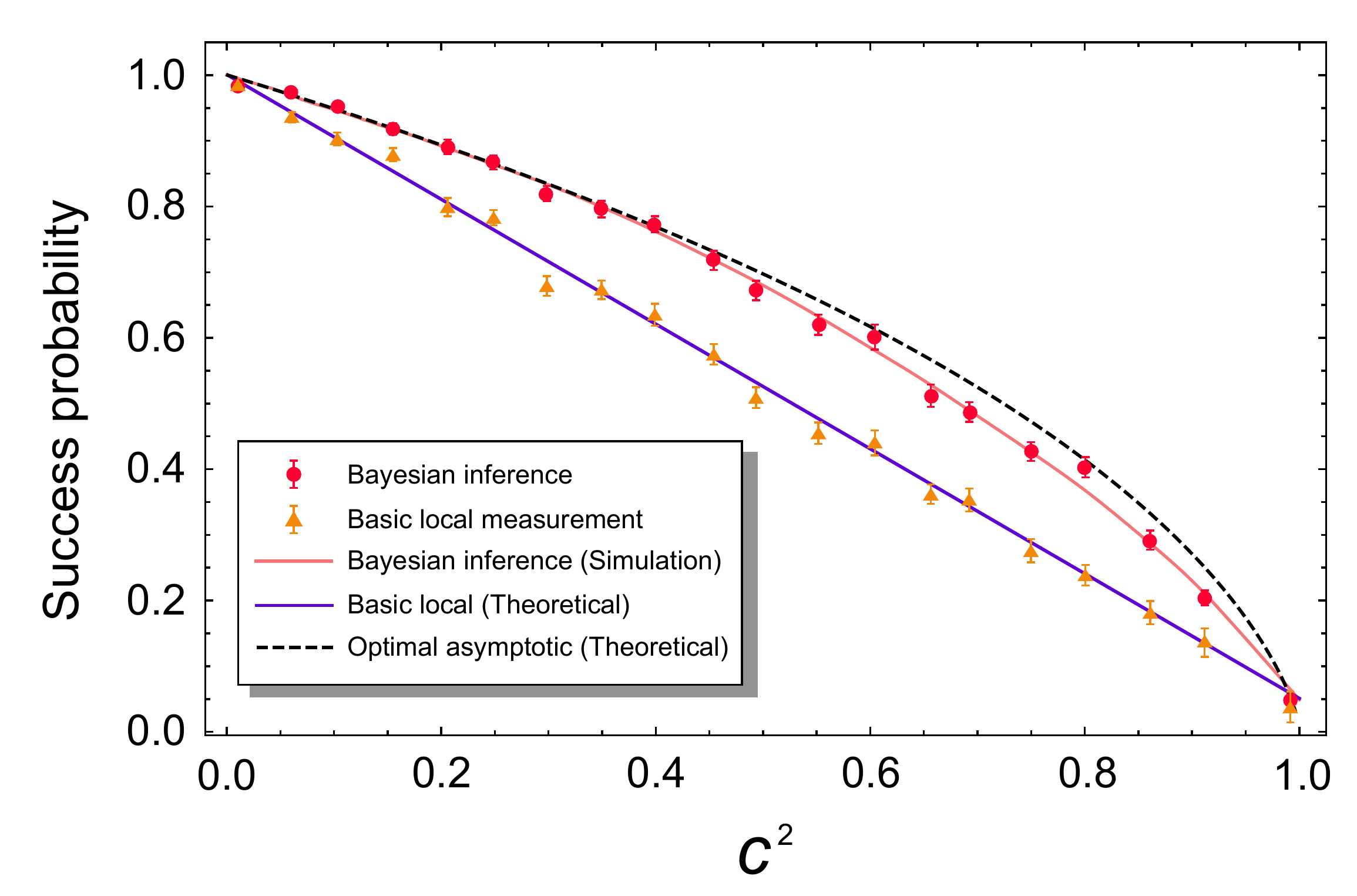}
\caption{\label{Fig4} Comparison of the BI, BL, and optimal measurement for every overlap. The red dots are the experimental results obtained using BI detection, and the solid light red line is the corresponding simulation value. The orange triangles are the results obtained from BL detection, and the solid purple line is its theoretical value. The success probability for the optimal measurement is shown as the dash black line. Although the BI approach cannot surpass the optimal (global) detection, it nearly covers the gap between the BL and the optimal measurement.}
\end{figure}

Finally, to assess the overall performance of each protocol, we conduct the aforementioned experiments and compute the corresponding average success probability for all possible change points and for each value of the overlap $c^2$, that we space in intervals of $\Delta c^2=0.05$. We show our results in Fig. 4. The red dots and orange triangles are the experimental results for the BI and BL strategies, respectively. The black dashed line represents the optimal measurement, and the purple solid line indicates the theoretical BL detection limit. It is clear from the results that the improvement over the BL strategy  provided by a machine-learning-enhanced local procedure  almost closes the gap with  the optimal measurement (which is extremely hard to implement, especially for large $n$).

Although the theoretical optimal detection strategy establishes an ultimate performance bound for the task of identifying quantum change points, it had yet to be seen how close could a realistic experimental implementation get. We have demonstrated that BI is an easily implementable local strategy, within the reach of current experimental techniques, that performs quasi-optimally. The identification of quantum change points demonstrated in our experiment might be applicable in many practical situations, such as identifying domain walls in condensed matter systems~\cite{Wang2015} and detecting the changes of fluorescence polarization in some biological processes~\cite{Hall2016}. Our work can also be extended to high dimensional quantum states, mixed states, and multiple change points. In addition, while our experimental setup uses a classical learning agent, the BI approach is also implementable with quantum learning algorithms~\cite{Low2014,Wiebe2015}. This will be studied in the future.

In summary, we report the first demonstration of the quantum change point detection based on the Bayesian inference strategy, which yields a higher success probability compared to other local methods. The BI approach contains a learning agent that can adjust the measurement basis at each step using Bayes rule. This agent gives a very significant advantage that yields a performance very close to optimality and also provides the capacity to correct mistakes caused by experimental noise.  The ability to accurately detect a quantum change point has an immediate impact on many quantum information tasks. Our learning method demonstrated that the protocol can be efficiently implemented to identify the position of a sudden change in quantum settings where a sequence of identical quantum states is required.

This work is supported by the National Key Research and Development Program of China (No. 2017YFA0304100), the National Natural Science Foundation of China (Grants Nos. 61327901, 11674304, 61490711, 11774334, 11774335, 11474267, 11325419, 11574291 and 91321313), the Key Research Program of Frontier Sciences of the Chinese Academy of Sciences (Grant No. QYZDY-SSW-SLH003), the Youth Innovation Promotion Association of Chinese Academy of Sciences (Grants No. 2017492), the Fundamental Research Funds for the Central Universities (No. WK2470000026). C.-F.L. acknowledges support from the EU Collaborative project QuProCS (641277). RMT acknowledges  the Spanish MINECO through contracts FIS2013-40627-P \& FIS2016-80681-P. GS acknowledges financial support from the ERC Consolidator Grant No. 683107/TempoQ, and the DFG.
~\\

S.Y. and C.-J.H. contributed equally to this work.

\bibliographystyle{}

\begin{thebibliography}{38}
\bibitem{results}
Here, ``0'' represnets the default state is detected and ``1'' means the mutated state is detected.
\bibitem{position}
The success probabilities when $k=4,12,18$ also reach the optimal measurement values, but we conclude it is caused by the experimental errors (see Appendix~\ref{app:E}), which means they are likely to fluctate below the optimal value in the next experiment.
\end{thebibliography}

\appendix 

\section{The BI protocol}\label{app:A}
The source state prepared by Alice can be expressed as $|\Psi_{k}\rangle=|H\rangle^{\otimes k-1}|\phi\rangle^{\otimes n-k+1}$, where the $k$ denotes the position of the change point and $n$ is the number of the photons in source state. Here, $\vert \phi \rangle=c\vert H \rangle+s\vert V \rangle$ ($s=\sqrt{1-c^{2}}$, and $H, V$ represent the horizontal and vertical polarization respectively). Without loss of generality, we choose $c$ to be real and positive.

\emph{Basic local}--- For a chain of photons which the polarization state begins with $|H\rangle$ and changes into the state $|\phi\rangle$ at an unknown point $k$, the simplest online strategy (namely, basic local) is to measure each photon in the basis $\{|H\rangle,|V\rangle\}$. The measurements are performed sequentially until the outcome $\vert \phi \rangle$ (i.e., get result ``1'') is obtained for the first time at the $r$th step. We are sure that the $r$th particle was in the state $|\phi\rangle$, which means that the change must have occurred at some position $k\leq r$. Then our best guess for the change point is $k=r$. The success probability is $p=1-c^2+\frac{c^2}{n}$.

We want to note here that the measurement basis need not be $\{|H\rangle,|V\rangle\}$. We also can set it as others, such as the Helstrom basis $\{\Pi(H),\Pi(\phi)\}$~\cite{Helstrom1976}, which is given by the projectors onto the positive and negative parts of the spectrum of the Helstrom matrix $\Gamma=\vert H\rangle\langle H \vert-\vert \phi\rangle\langle \phi \vert$. However, it can be proven that the success probability of all the fixed-basis methods cannot surpass the learning-enhanced local strategy~\cite{preparation}.

\emph{Bayesian inference}--- This online strategy can improve the success probability by adjusting the measurement basis according to the experimental result in each step. We build a classical learning agent to guess where the change point occurs. It starts with a uniform prior $p(k)=\frac{1}{n}$ about the hypothesis of the change point and updates the expectation as new data obtained. In order to update the information at the $s$th step, the learning algorithm is designed as follows~\cite{Sentis2016}:

\begin{enumerate}
\item Find the probability $p^{(s)}_H$ (or $p^{(s)}_\phi$) of the most likely sequence that has the particle at position $s$ being in the state $|H\rangle$ (or $|\phi\rangle$), which in general depends on all previous results $r_1,r_2,\cdots,r_{s-1}$ through the priors $\eta^{(s)}_{k}\equiv p(k|r_1,r_2,\cdots,r_{s-1})$:
$$p^{(s)}_H=\max\limits_{k}\{\eta^{(s)}_{k}\}^n_{k=s+1}$$
$$p^{(s)}_\phi=\max\limits_{k}\{\eta^{(s)}_{k}\}^s_{k=1}$$
\item Perform the Helstrom measurement~\cite{Helstrom1976} on the $s$th photon. The measurement basis $\{\Pi_{s}(H),\Pi_{s}(\phi)\}$ is given by the projectors onto the positive and negative parts of the spectrum of the Helstrom matrix $\Gamma_{s}=p_{H}^{(s)}\vert H\rangle\langle H \vert-p_{\phi}^{(s)}\vert \phi\rangle\langle \phi \vert$
\item After the $s$th measurement has been performed, the prior is updated in accordance with the measurement result, using Bayes' update rule:
$$\eta^{(s+1)}_k=\frac{p(r_s|k)\eta^{(s)}_k}{\sum^n_{l=1}p(r_s|l)\eta^{(s)}_l}$$
\item Go back to the first procedure for next measurement.
\end{enumerate}
After the last measurement, the agent updates the prior to $\eta^{(n+1)}_k$ and produces the guess $\hat{k}$ that maximizes $\eta^{(n+1)}_{k}$ for the change point ($\hat{k}=\text{argmax}_{k}(\eta^{(n+1)}_{k})$). If the guess is correct ($\hat{k}=k$), mark this experiment as successful. Otherwise, mark it as unsuccessful.

\section{The logic unit postselection}\label{app:B}
The particles in the source state $\vert\Psi_{k}\rangle$ come from a heralded single-photon source based on the spontaneous parametric down conversion (SPDC) process in a 1-mm-thick beta barium borate (BBO) crystal in our experiment, and the coincidence window is 3 ns. Here, we apply the shorthand notation $e_{t}^{\xi}$ to represent a event collected by the detector at time $t$, and $\xi\in\{\text{trigger},\text{signal},\text{idler}\}$ (signal=H or V). Meanwhile, we define that $\text{Coin}(e_{t}^{\xi},e_{t'}^{\xi'})=\text{true}$, if $\vert t-t'\vert <3 \, \text{ns}$.

Now, we define the herald single photon. If
\begin{equation*}\label{hsp}
\begin{aligned}
\text{Coin}(e_{t_{0}}^{\text{idler}},e_{t_{0}'}^{\text{signal}})=\text{true},
\end{aligned}
\end{equation*}
we call $e_{t_{0}'}^{\text{signal}}$ is a herald single photon event, which is detected at the transmission path (second port) or the reflex path (third port), shown in Fig. 2(a).

Because background noise exists in the experiment, we need to collect the true signals that come from SPDC process, namely, the effective event. The effective event is defined as two cases. \\
Case 1: If \\
\begin{equation*}\label{c1}
\begin{aligned}
& \text{Coin}(e_{t_{0}}^{\text{idler}},e_{t_{0}'}^{\text{H}})=\text{true}, \quad \text{and}\\
& \text{Coin}(e_{t_{0}}^{\text{idler}},e_{t_{0}'}^{\text{V}})=\text{false},
\end{aligned}
\end{equation*}
we call it as an effective event that $\vert H \rangle$ is detected.\\
Case 2: If \\
\begin{equation*}\label{c2}
\begin{aligned}
& \text{Coin}(e_{t_{0}}^{\text{idler}},e_{t_{0}'}^{\text{V}})=\text{true}, \quad \text{and}\\
& \text{Coin}(e_{t_{0}}^{\text{idler}},e_{t_{0}'}^{\text{H}})=\text{false},
\end{aligned}
\end{equation*}
we call it as an effective event that $\vert \phi \rangle$ is detected.

Then, we postselect the first effective event in each bin by using the trigger signal and an appropriate delay $\Delta t$. At the $s$th step, we set $\Delta t \in(T,T+2.5\, \text{ms})$ and $T=(s-1)\times 5\, \text{ms}$ (we need to detect the $s$th particle in the source state).

\, If \\
\begin{equation*}\label{ps}
\begin{aligned}
& \text{Coin}(e_{t+\Delta t}^{\text{trigger}},e_{t_{0}}^{\text{signal}})=\text{true}, \quad \text{and} \\
& \text{satisfy} \qquad \text{Case 1} \quad or \quad \text{Case 2}
\end{aligned}
\end{equation*}
we select the earliest detected event $e_{t_{0}}^{\text{signal}}$ as the first effective event in the $s$th time bin. In this situation, if Case 1 is satisfied, we regard it as we detect a default state, i.e., we obtain result ``0'' in this measurement. Conversely, if Case 2 is satisfied, we regard it as the photon that we detected suffers a polarization rotation, i.e., a mutated state is detected and we obtain result ``1'' in this measurement.

\section{Numerical simulations for different values of $n$ and different change points}\label{app:C}
In our experiments, we find that the success probability has a certain relationship with the position of the change point $k$ (shown in Fig. 3(b)). Therefore, we do some numerical experiments to study its relationship for different $n$ (shown in Fig. A1). As we can find in Fig. A1(a-e), the success probabilities are much higher than others when $k=1$ or \emph{n} and all of them are better than basic-local one. From Fig. A1(f), we can conclude the fact that although the success probabilities decrease when $n$ increases, the differences of successful probability between the Bayesian inference and basic-local method is nearly unchanged, which means the Bayesian inference strategy indeed has advantage over the basic local strategy, regardless the number of photons $n$.
\begin{figure*}[ht]
\centering
\includegraphics[width=1.0\textwidth]{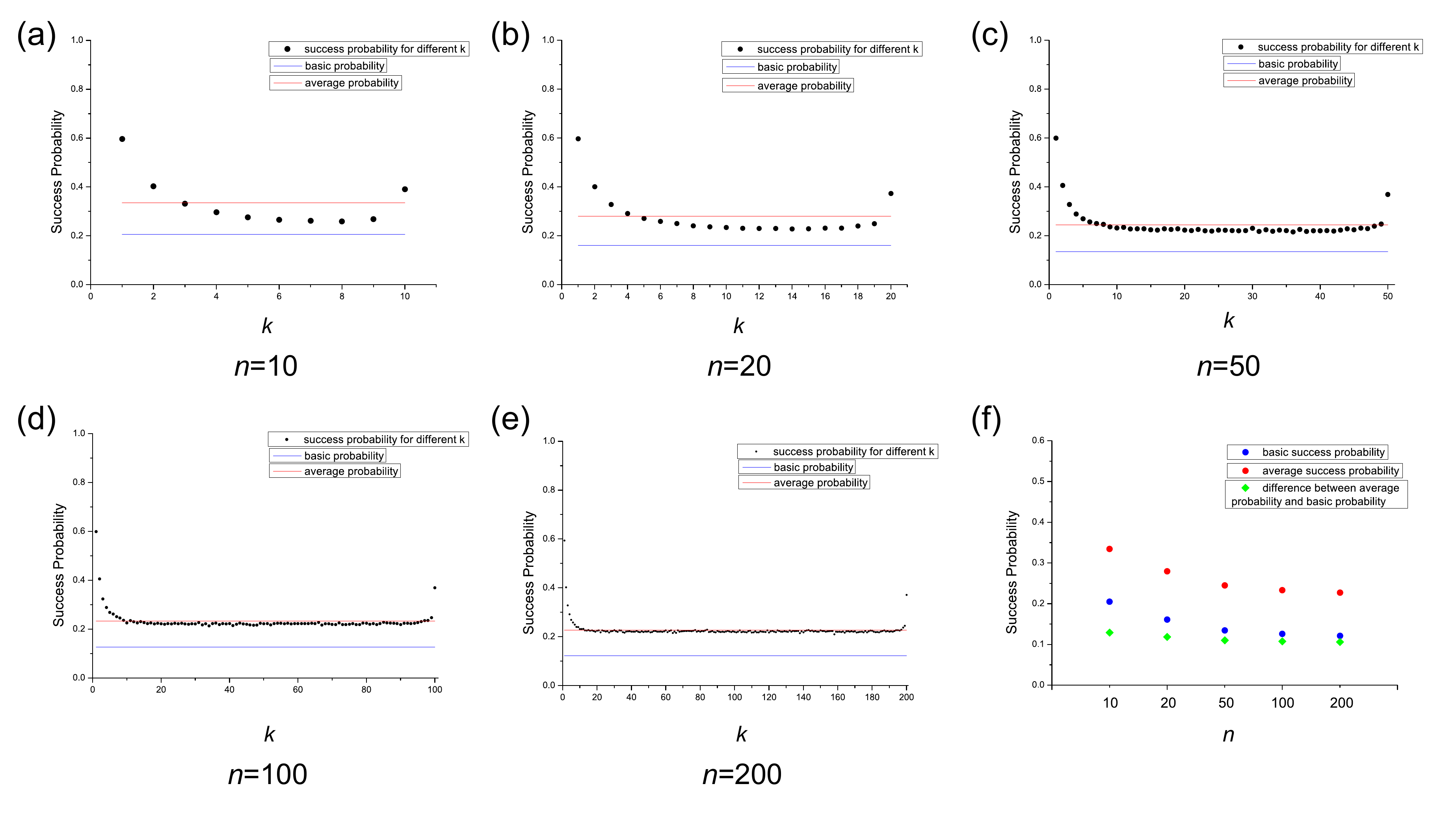}
\renewcommand{\thefigure}{A1}
\caption{\label{FigA1} The numerical results for different $n$. For Fig. A1(a-e), The red line stands for average success probability (BI strategy). The blue line stands for the success probability of the BL strategy. The black dots stand for success probability for different $k$ (BI strategy). Here we set $c^{2}=0.883$. For Fig. A1(a-b), we take 100,000 numerical experiments for each point. For Fig. A1(c-e), we take 20,000 numerical experiments for each point. Fig. A1(f) shows the compare between different $n$. The red dots stand for average success probability (BI strategy). The blue dots stand for the success probability of the BL strategy. The green dots stand for the difference between the average BI and BL success probability.}
\end{figure*}

\section{Advantage of Bayesian inference approach}\label{app:D}
In order to find a more clear advantage of Bayesian inference approach, we plot the distances between the BI approach with respect to the basic local and the optimal measurement (shown in Fig. A2). The red bars are the improvement of BI over the BL detection strategy, and the blue bars are the distance between BI and the optimal measurement.
\begin{figure}[tb]
\centering
\includegraphics[width=0.5\textwidth]{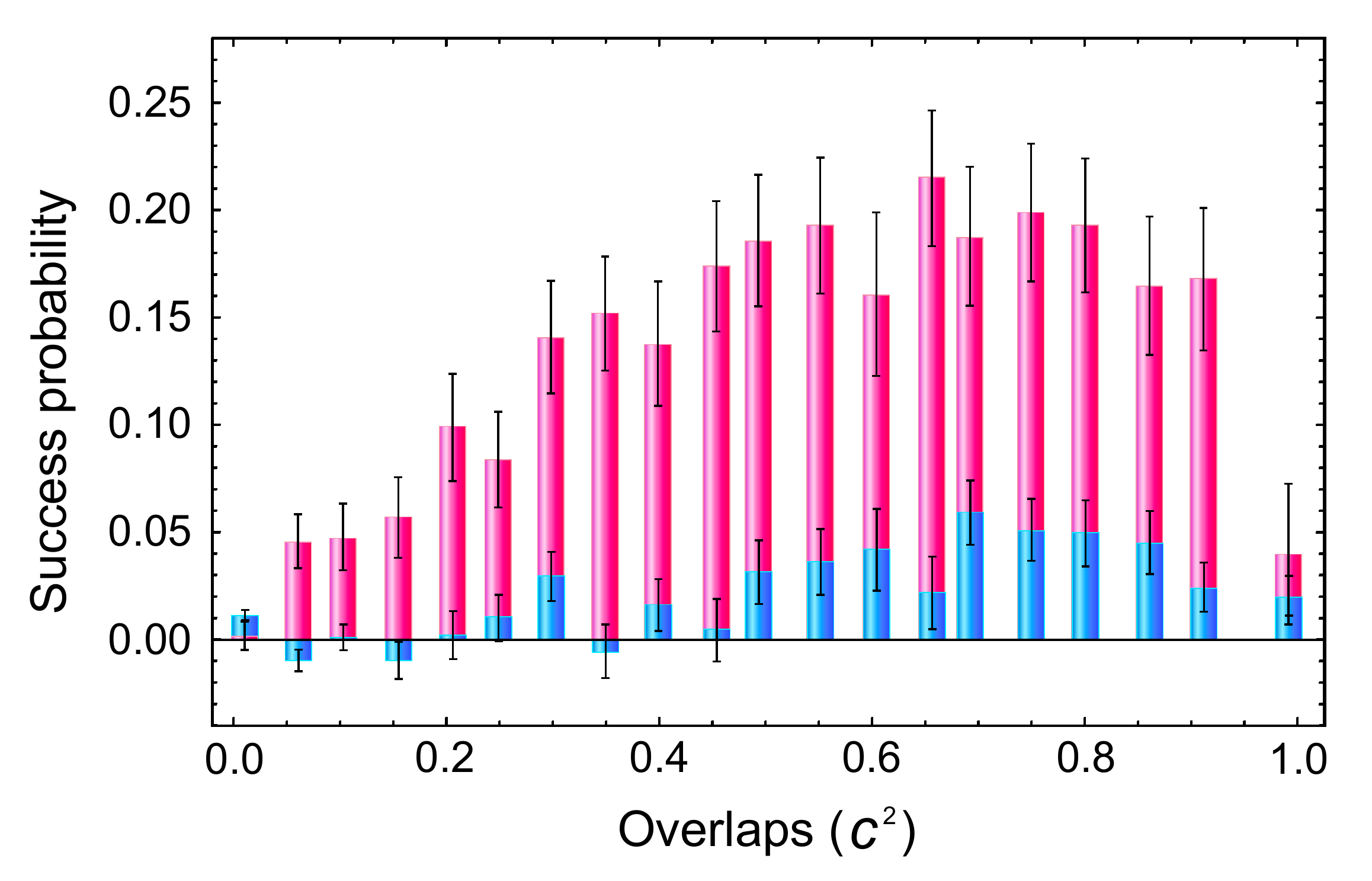}
\renewcommand{\thefigure}{A2}
\caption{\label{FigS2} The advantage of the Bayesian inference strategy. The red bars are the improvement of BI with respect to the BL detection strategy, and the blue bars are the distance between BI and the optimal measurement.}
\end{figure}

\section{Experimental error analysis}\label{app:E}
In our experiment, the experimental error is caused by several aspects, such as the background noise, the systematic error and the statistical error. In Fig. 4, the success probabilities at $c^{2}=0.010$ for two strategies are $P_{\text{BI}}=0.984\pm0.002$ and $P_{\text{BL}}=0.982\pm0.004$, respectively.

Because of the imperfect experimental apparatuses, such as the EOM and the PBS, we cannot obtain an extremely high extinction ratio. In other words, we cannot create the ideal orthogonal state ($c^{2}=0$). In this case ($c^{2}=0.010$), the default state and the mutated state are nearly orthogonal, and the statistical error is minimized since all the experimental results in this case should give the correct guessing point. Therefore, the distances between the experimental data and the corresponding theoretical values ($P^{th}_{\text{BI}}=0.995$ and $P^{th}_{\text{BL}}=0.991$) are mainly caused by the background noise and the systematic error. The fluctuation of success probabilities that appears at other places is mainly caused by the statistical error. Since we run 50 experiments for each change point at one overlap, the statistical probability obtained by 50 experiments (Fig. 3(b)) and 1000 experiments (Fig. 4) will have a statistical fluctuation. We can find a relatively large fluctuation (a relative large error-bars) in Fig. 3(b) because there are just 50 experiments for each change point. However, in Fig. 4, the fluctuation (or error-bars) becomes much smaller since they are calculated from totally 1000 experiments (50 times for each change point and there are 20 possible change positions for each overlap). Although the number of experiments for each source state is not very large, the experimental results for different $k$ (or $c^{2}$) matches the numerical results and shows the characteristic of the relationship between success probability and the change position (or overlap).

\end{document}